# Comparing Unloaded Q-Factor of a High-Q Dielectric Resonator Measured Using The Transmission Mode and Reflection Mode Methods Involving S-Parameter Circle Fitting.

K. Leong[*], J. Mazierska, M.V. Jacob, D. Ledenyov, S. Batt

Electrical and Computer Engineering, James Cook University, Townsville. 4811. Australia
* now with the National Institute of Standards and Technology, Boulder, CO 80305-3328, USA

*Abstract* — **A comparative study of unloaded Q-factor measurements of a $TE_{011}$ mode sapphire dielectric resonator with unloaded Q-factor value of 731,000 at a frequency of 10 GHz and temperature of 65 K using two best Q-factor measurement methods are presented. The Transmission (TMQF) and Reflection methods are based on relevant multifrequency S-parameter measurements and circle-fitting procedures to compute the unloaded Q-factor of the resonator. For accurate comparison of the methods a delay compensation procedure (introduced in the TMQF technique to remove delay due to noncalibrated cables) has been applied also to the reflection data.**

## I. INTRODUCTION

Accurate measurements of microwave properties of materials using dielectric resonators require precise computations of the unloaded Q factor of the resonator. There are various methods which enables measurement the Q-factors of microwave resonators [1-14]. However, not all of them take into account practical effects introduced by a real measurement system. The practical effects include noise, crosstalk, coupling losses, transmission line delay, and impedance mismatch. Inadequate accounting of the practical effects may lead to significant uncertainty in the Q-factor obtained [8,13,14,15].

Two of the most accurate and practical Q-factor methods used today include the Reflection method for reflection mode resonators [1] and the Transmission Mode Quality Factor Technique for transmission mode resonators [13,15]. The methods involve circle-fitting procedures applied to multiple data points representing S-parameter responses of the resonator around the resonance. The circle fitting techniques have become a popular choice for determination Q-factors of dielectric resonators because measurements of S-parameters can be done accurately and conveniently using vector network analysers.

This paper presents results of the accuracy comparison of the Reflection method and the TMQF technique.

## II. TRANSMISSION AND REFLECTION MODE METHODS FOR COMPUTATION OF THE Q-FACTOR

The Transmission Mode Q-Factor (TMQF) method and the Reflection Mode method applicable to resonators working in the transmission mode and the reflection mode are based on the circuit models of a dielectric resonator system shown in Fig. 1 and Fig. 2 respectively.

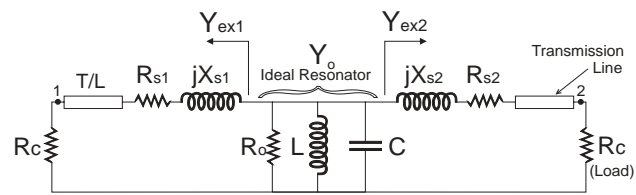

Fig. 1 Circuit model of a transmission mode dielectric resonator system used to develop the Transmission Mode Q-Factor technique [13].

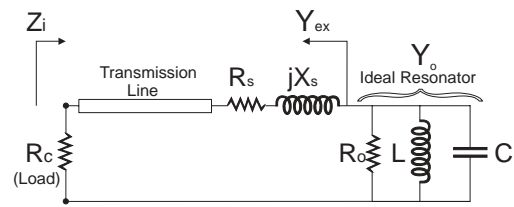

Fig. 2 Circuit model of a reflection mode dielectric resonator system used to develop the Reflection Mode method[1].

Each of the circuits shows an ideal resonator connected to a external circuit, and incorporates elements to account for a lossy coupling and coupling reactance ($R_s$ and $X_s$). The fundamental equations to calculate the unloaded $Q_o$-factor from the loaded $Q_L$-factor and the coupling coefficient(s) for the transmission and reflection mode resonators are:

$$Q_o = Q_L(1 + \beta_1 + \beta_2) \quad (1)$$

and

$$Q_o = Q_L(1 + \beta) \quad (2)$$

The loaded $Q_L$-factor and the coupling coefficients are determined from the circle fitting procedures of the Q-factor methods, which requires measurements of S-parameter response(s) of the resonator around the resonance. Relationships describing $S_{21}$, $S_{11}$, and $S_{22}$ with $Q_L$, $\beta_1$ and $\beta_2$ for the TMQF technique are given below after [13,15]:

$$S_{21}(\omega) = \frac{2R_c Y_{ex1} Y_{ex2}}{G_o(1+\beta_1+\beta_2)\left[1+j2Q_L\frac{(\omega-\omega_L)}{\omega_o}\right]}$$

$$\approx \frac{2R_c Y_{ex1} Y_{ex2}}{G_o(1+\beta_1+\beta_2)\left[1+j2Q_L\frac{(\omega-\omega_L)}{\omega_L}\right]} \quad (3)$$

where $\omega_o$ and $\omega_L$ are the unloaded and loaded resonant frequencies[13]. For the Reflection Mode method [1], the fundamental equation relating $S_{11}$ with $Q_L$ and $\beta$ is:

$$S_{11} = \frac{S_{11d} + de^{j\gamma} + j2Q_L S_{11d}\left(\frac{\omega-\omega_L}{\omega_o}\right)}{1+jQ_L 2\left(\frac{\omega-\omega_L}{\omega_o}\right)} \quad (4)$$

where $d = d_2\frac{\beta}{1+\beta}$, $d_2$ is a complex constant, $S_{11d}$ is the detuned value of $S_{11}$ at frequencies far from the resonant frequency, and

$$\gamma = -2\tan^{-1}\left[\frac{X_s}{R_c + R_s}\right] \quad (5)$$

For transmission ($S_{21}$) and reflection ($S_{11}$ or $S_{22}$) parameters of the dielectric resonator system around resonance, the path of the S-parameter vector traced in the complex plane is ideally a circle referred as a 'Q-circle'. The Q-circle for all three S parameters has the functional form [1]:

$$S = \frac{a_1 t + a_2}{a_3 t + 1} \quad (6)$$

where $a_1$, $a_2$ and $a_3$ are three complex constants, and 't' is a normalised frequency variable defined by the following function of frequency:

$$t = 2\left(\frac{\omega-\omega_L}{\omega_L}\right) \quad (7)$$

The constants $a_1$, $a_2$ and $a_3$ are computed using the Fractional Linear Curve Fitting Technique applied to N points measured around the resonance [1].

In both the transmission mode and reflection mode methods, the loaded $Q_L$-factor is simply:

$$Q_L = \text{Im}[a_3] \quad (8)$$

However, for the transmission mode method $Q_L$-factor is obtained from the circle-fitting to the $S_{21}$ Q-circle, while the reflection mode method it is obtained from the $S_{11}$ Q-circle.

The coupling coefficient $\beta$ for the reflection mode resonator [1] is considered to be a sum of lossless $\beta_i$ and lossy parts $\beta_L$:

$$\beta = \beta_i + \beta_L \quad (9)$$

where

$$\beta_i = \frac{1}{2\left(\frac{1}{d_Q} + \frac{1}{d_L}\right)}, \quad \beta_L = \frac{\left(\frac{2}{d_L}-1\right)}{2\left(\frac{1}{d_Q}+\frac{1}{d_L}\right)} \quad (10)$$

The total port coupling coefficient is simply computed from the respective diameters $d_Q$ and $d_L$ of the Q-circle and coupling loss circle as:

$$\beta = \frac{1}{\frac{d_L}{d_Q}-1} \quad (11)$$

The coupling loss circle shown in Fig. 3 is defined as the circle that lies tangential to the-reflection $S_{11}$ Q-circle and the unit circle [1]. The diameters of the Q-circle and the coupling loss circle are determined using equations (12) and (13) respectively:

$$\text{diameter of } Q \text{ circle} = \left|a_2 - \frac{a_1}{a_3}\right| \quad (12)$$

$$\text{Diameter of Loss Circle}\Big|_{port\,p} = \frac{1-|S_{ppd}|^2}{1-|S_{ppd}|\cos\phi} \quad (13)$$

where the detuned value $S_{ppd}$ is the value of the S-parameter at frequencies far away from the resonant frequency.

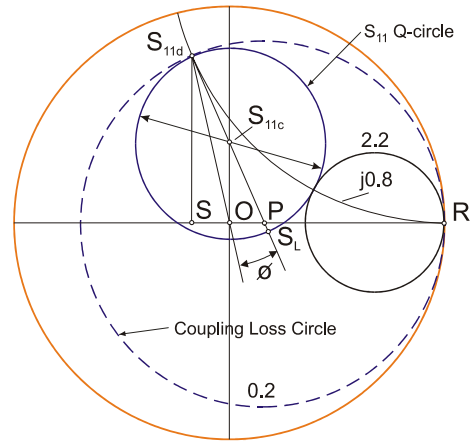

Fig. 3 Q-circle and Coupling Loss circle based on [1]

For the transmission mode resonator, there are two coupling coefficients and they have been considered in the same way as in [1] to consist of the lossless and lossy parts.

$$\beta_p = \beta_{pi} + \beta_{pL} \quad (14)$$

where the port number is denoted by p (= 1 or 2).
The equations describing the lossless $\beta_{pi}$ and lossy $\beta_{pL}$ parts of the port coupling coefficient are different than that for the reflection method, and these have been obtained in [13,15] as:

$$\beta_{pi} = \frac{x}{2\left[1-\left(\frac{x}{d_1}+\frac{y}{d_2}\right)\right]}, \quad \beta_{pL} = \left(\frac{2}{d_p}-1\right)\beta_{pi} \quad (15)$$

where $d_1$ is a diameter of port 1 coupling loss circle, $d_2$ is a diameter of port 2 coupling loss circle, x is a diameter of port 1 ($S_{11}$) Q-circle, y is a diameter of port 2 ($S_{22}$) Q-circle. The diameters of the $S_{11}$ and $S_{22}$ Q-circles and coupling loss circles for the transmission mode method are determined using the same equations (12) and (13) as for the reflection mode method.

III MEASUREMENTS OF $Q_o$ FACTOR OF THE SAPPHIRE DIELECTRIC RESONATOR USING THE TRANSMISSION MODE Q-FACTOR METHOD AND THE REFLECTION METHOD

The measurement system shown in Fig 4 includes HP8722C vector network analyser, vacuum dewar, a temperature controller, and a Hakki-Coleman sapphire dielectric resonator with copper cavity and superconducting end-walls.

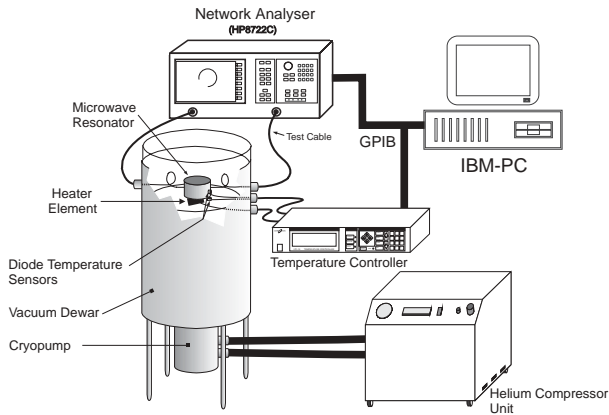

Fig. 4   Measurement system used in the comparison measurements of unloaded $Q_o$-factor of the sapphire dielectric resonator.

The resonator is coupled to the external system (on each side) via a coupling loop situated at the end of a semi-rigid cable as shown in Fig. 5. The other end of the cable connects to a feed-through connector on the wall of the dewar. Since the dielectric resonator used in the tests is essentially a transmission mode resonator, measurements for Reflection Mode resonator tests were made on port 1 side while the port 2 side was fully decoupled.

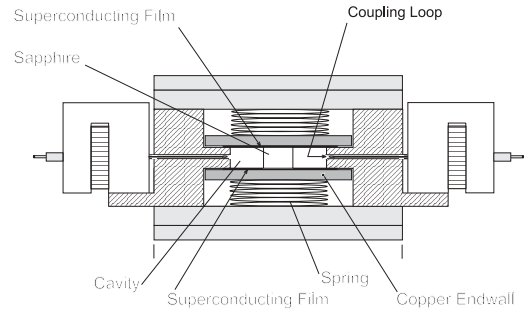

Fig. 5 The $TE_{011}$ mode sapphire dielectric resonator used for the $Q_o$-factor comparison tests.

The comparison test involved using each technique to determine the $Q_o$-factor of the dielectric resonator for different levels of coupling. The transmission mode tests required measurements of three S-parameters ($S_{21}$, $S_{11}$, and $S_{22}$) for each level of coupling. For the reflection mode test, only one parameter ($S_{11}$) was needed. As mentioned in the Abstract a procedure to remove the phase delay developed for the TMQF technique has been applied to the Reflection technique.

Results of the $Q_o$-factor comparison test for different positions of the tips of the coupling loop with respect to the lateral wall of the cavity are shown in Fig. 6.

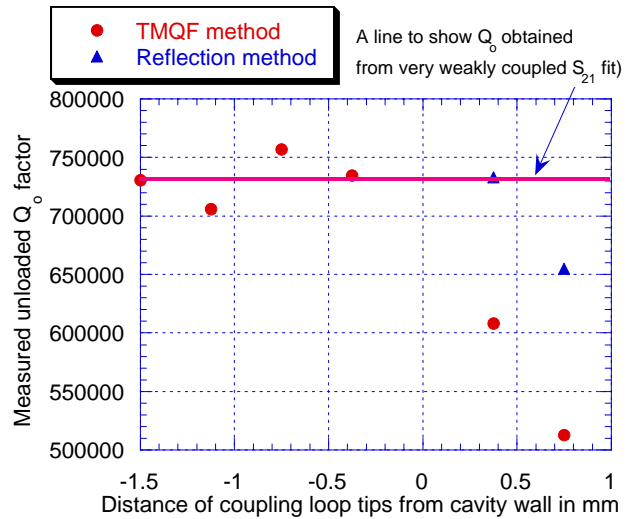

Fig. 6 Measured $Q_o$-factor of the sapphire resonator using the Transmission (TMQF) and Reflection techniques.

The positions of the loop tips range from 1.5 mm external to the cavity to 0.75 mm internal to the cavity (the negative values of loop distance in the figure indicates positions exterior to the cavity). The value of the $Q_o$-factor obtained for the very weak coupling of 730742 was used as a reference and shown as the line in Fig. 6. Values of coupling coefficients for varying loop positions are listed in Table 1.

TABLE 1. Coupling coefficients obtained for various loop positions.

| Loop distance in mm | $\beta_1$, $\beta_2$ (TMQF) | $\beta$ (Reflection method) |
|---|---|---|
| 0.750 | 1.652, 1.381 | 2.064 |
| 0.375 | 0.919, 0.806 | 0.996 |
| -0.375 | 0.138, 0.090 | - |
| -0.750 | 0.059, 0.041 | - |

## IV. CONCLUSIONS

Results of a comparison study of unloaded Q-factor measurements of two best Q-factor determination methods have been presented, namely the Transmission Mode and Reflection Mode methods. Under the test conditions described in III, the results show that the accuracy of the two methods are comparable. However the TMQF technique enables measurements with small coupling coefficients what is not feasible for the Reflection technique. The error in $Q_o$-factors is estimated to be less than 3% for loop tip distances between –0.4 mm to +0.4 mm corresponding to port coupling coefficient values ranging from about 0.4 to 1.0. For these levels of coupling, the observed reflection Q-circles were well-defined with signal-to-noise ratio greater than 27.5 dB.

For coupling loop positions larger than 0.4 mm inside the cavity, a significant reduction in the $Q_o$-factor obtained from both Q-factor methods is observed. The authors expect the observed effect is due the presence of the coupling loops and semi-rigid cable inside the cavity, which reduces the unloaded $Q_o$-factor of the system.

For loop positions larger than 0.4 mm outside the cavity, the coupling becomes so small that reflection $S_{11}$ and $S_{22}$ Q-circles cannot be measured accurately. As coupling is gradually reduced, it has been observed that measurements of reflection ($S_{11}$ and $S_{22}$) Q-circles become unreliable well before the same problems with $S_{21}$ Q-circles occur following [13]. Hence it is expected that the accuracy of the Transmission Mode method can provide higher accuracy for very weak coupling conditions, but not so weak to encounter unreliable measurements of $S_{21}$ Q-circles.

**Acknowledgement:** This work was done under the ARC Large Grant A079 (James Cook University) what is gratefully acknowledged.